\documentclass[12pt]{article}

\usepackage{color}
\usepackage{graphicx}

\usepackage{times}

\newenvironment{sciabstract}{%
\begin{quote} \bf}
{\end{quote}}

\newcounter{lastnote}

\title{Giant phonon anomalies and central peak due to charge density wave formation in YBa$_2$Cu$_3$O$_{6.6}$}

\author{M. Le Tacon$^{1, \ast}$, A. Bosak$^{2}$, S. M. Souliou$^{1}$, G. Dellea$^{3}$, T. Loew$^{1}$\\
R. Heid$^{4}$, K.-P. Bohnen$^{4}$, G. Ghiringhelli$^{3}$, M. Krisch$^{2}$, B. Keimer$^{1, \ast}$\\
\\
\normalsize{$^{1}$ Max-Planck-Institut~f\"{u}r~Festk\"{o}rperforschung,}\\
\normalsize{Heisenbergstra{\ss}e 1, D-70569 Stuttgart, Germany}\\
\normalsize{$^{2}$ European Synchrotron Radiation Facility, BP 220, F-38043 Grenoble Cedex, France}\\
\normalsize{$^{3}$ CNR-SPIN, CNISM and Dipartimento di Fisica,}\\
\normalsize{Politecnico di Milano, Piazza Leonardo da Vinci 32, I-20133 Milano, Italy}\\
\normalsize{$^{4}$ Institut f\"{u}r Festk\"{o}rperphysik, Karlsruher Institut f\"{u}r Technologie (KIT),}\\
\normalsize{P.O.B. 3640, D-76021 Karlsruhe, Germany}\\
\\
\normalsize{$^\ast$To whom correspondence should be addressed;}\\
\normalsize{E-mail: m.letacon@fkf.mpg.de, b.keimer@fkf.mpg.de.}
}

\date{}

\begin{document}


\baselineskip24pt


\maketitle


\begin{sciabstract}
The electron-phonon interaction is a major factor influencing the competition between collective instabilities in correlated-electron materials, but its role in driving high-temperature superconductivity in the cuprates remains poorly understood. We have used high-resolution inelastic x-ray scattering to monitor low-energy phonons in  YBa$_2$Cu$_3$O$_{6.6}$ (superconducting $\bf T_c = 61$ K), which is close to a charge density wave (CDW) instability. Phonons in a narrow range of momentum space around the CDW ordering vector exhibit extremely large superconductivity-induced lineshape renormalizations. These results imply that the electron-phonon interaction has sufficient strength to generate various anomalies in electronic spectra, but does not contribute significantly to Cooper pairing. In addition, a quasi-elastic ``central peak'' due to CDW nanodomains is observed in a wide temperature range above and below $\bf T_c$, suggesting that the gradual onset of a spatially inhomogeneous CDW domain state with decreasing temperature is a generic feature of the underdoped cuprates.
\end{sciabstract}

The character of the electron-phonon interaction (EPI) in metals and superconductors with strongly correlated electrons has recently been the subject of intense research, especially with regard to its influence on high-temperature superconductivity in the copper oxides \cite{Gunnarsson,Capone,Reznik1}. Whereas density-functional theory (DFT) predicts only weak EPIs in these materials \cite{Savrasov,Bohnen,Giustino,Heid}, various anomalies in the dispersion relations of both conduction electrons and lattice vibrations have been interpreted as evidence of an EPI strength far exceeding these predictions. In particular, the energies of prominent ``kinks'' in the electronic band dispersions \cite{Cuk,Devereaux,Vishik,Johnston} were reported to be in good agreement with those of Cu-O bond vibrations that also exhibit sizable dispersion anomalies \cite{Pintschovius,Reznik2,Raichle}. However, it has proven difficult to disentangle the influence of the EPI on these anomalies from other factors including spin fluctuations and lattice anharmonicity, respectively. The question whether strong correlations substantially modify the EPI in the cuprates with respect to standard DFT therefore remains open, as do more general questions about the role of the EPI in driving high-temperature superconductivity and/or competing instabilities such as ``stripe'' or charge density wave (CDW) order. Similar questions are being asked for other correlated metals including the recently discovered iron-based superconductors.

We approached these issues from a new direction, following leads from recent x-ray scattering experiments that revealed strong CDW correlations in underdoped YBa$_2$Cu$_3$O$_{6+x}$, one of the most widely studied high-temperature superconductors \cite{Ghiringhelli,Chang,Achkar,Blackburn,Blanco-Canosa}. Specifically, the CDW was shown to be present throughout the bulk of the material, clearly separated from magnetically ordered states, and competing strongly with superconductivity. YBa$_2$Cu$_3$O$_{6+x}$ is therefore well suited as a platform for the investigation of the role of the EPI as a mediator of collective ordering phenomena in the cuprates. However, the x-ray experiments reported thus far were carried out either in energy-integrating mode, or with an energy resolution insufficient to discriminate static and dynamic CDW correlations. Individual phonon modes could not be resolved in any of these experiments.

With this motivation, we have used non-resonant inelastic x-ray scattering (IXS) with high energy resolution to carefully monitor the temperature dependence of low-energy lattice vibrations around the CDW ordering wavevector of underdoped YBa$_2$Cu$_3$O$_{6.6}$. We observed pronounced phonon anomalies at energies that correspond well to those of recently discovered low-energy ``kinks'' in the electronic bands of cuprates at similar doping levels~\cite{Vishik,Johnston}. These anomalies confirm a large EPI that is, however, highly anisotropic and confined to a very narrow window of momentum space. In addition, we discovered a quasielastic ``central peak'' whose intensity is maximal at the superconducting transition temperature, $T_c$, but persists over a wide temperature range above $T_c$. In analogy to classical work on structural phase transitions \cite{Cowley,Axe}, we interpret this observation as evidence of a spatially inhomogeneous state in which lattice defects nucleate CDW nanodomains. These results have profound implications for the long-standing debate about the nature of electronic phase transitions in the ``pseudogap'' regime of the cuprate phase diagram.

\begin{figure}
    \begin{center}
        \includegraphics[width=0.9\columnwidth]{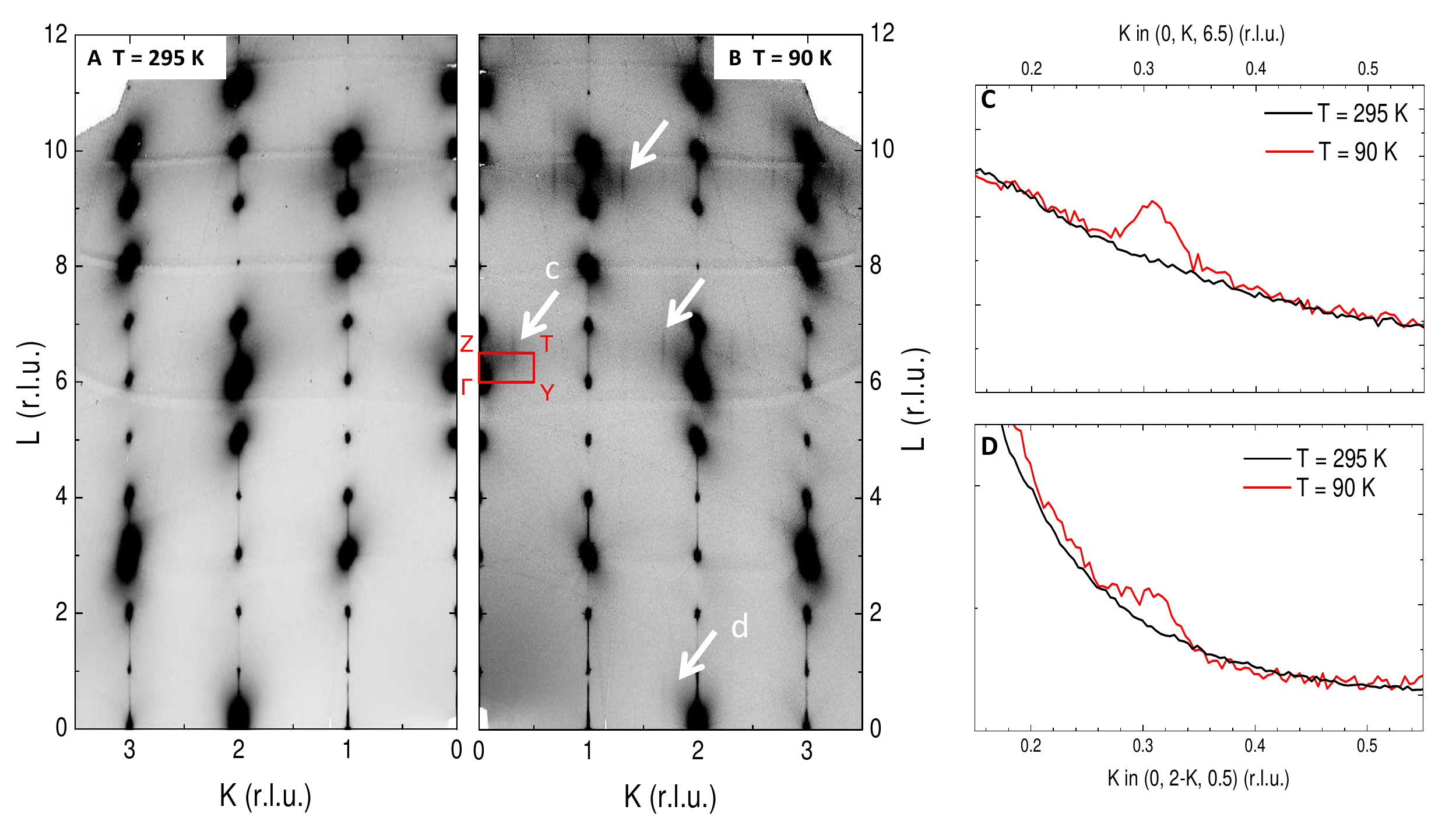}
    \end{center}
    \caption{Diffuse scattering mapping of the ${\bf Q} = (0,K,L)$ plane (A) at room temperature and (B) at $T =90$ K. White arrows indicate the CDW superstructure peaks. Panels (C) and (D) display cuts of these maps along $(0,K,6.5)$ and $(0, 2-K, 0.5)$, respectively. The red box corners correspond to the $\Gamma$, $Y$, $T$ and $Z$ points of the BZ centered at ${\bf Q} = (0,0,6)$.}
    \label{fig:DS}
\end{figure}

The IXS experiments were performed on an underdoped YBa$_2$Cu$_3$O$_{6.6}$ single crystal with $T_c = 61$ K previously studied with soft x-rays~\cite{Ghiringhelli,Blanco-Canosa}. In such experiments, the phonon intensity depends on the total momentum transfer $\bf Q$, rather than the reduced momentum $\bf q = Q-G_{HKL}$, where $\bf G_{HKL}$ stands for the Brillouin zone (BZ) center, $\Gamma$, closest to $\bf Q$. As high-resolution IXS measurements are performed with high-energy ($\sim 20$ keV) photons, the Ewald sphere encompasses hundreds of BZs. Prior to the IXS experiments, we therefore assembled a comprehensive map of the diffuse scattering intensity to identify the BZs with the most intense x-ray signatures of CDW formation. This method has recently been applied successfully to the study of low-energy phonons in other compounds including 2H-NbS$_2$~\cite{Leroux} and ZrTe$_3$~\cite{Hoesch}. In YBa$_2$Cu$_3$O$_{6.6}$, superstructures of oxygen dopants~\cite{Andersen,Strempfer} give rise to intense diffuse features in the $(H,0,L)$ plane of reciprocal space \cite{SOM}, which obscure the CDW reflections. We therefore focused on the $(0,K,L)$ plane, where the intensity maps (Fig.~\ref{fig:DS}A,B) reveal the emergence of extended features at low temperatures. The positions, ${\bf q}_{CDW} = (0,0.31,0.5)$, and correlation lengths of these features (Fig.~\ref{fig:DS}C,D) are compatible with those of the CDW reflections reported previously on the same material \cite{Ghiringhelli,Chang}. Their structure factor depends strongly on both $K$ and $L$, mirroring the complex ionic displacement pattern associated with CDW formation. In the following, we will focus on transverse acoustic and optical phonons in the BZ adjacent to $\bf G_{006}$, where the CDW features are particularly intense. Similar results were obtained for longitudinal phonons near $\bf G_{020}$ \cite{SOM}.

Figure 2 shows a survey of the phonon dispersions and IXS intensity in this BZ at room temperature, along with the results of DFT calculations in several high-symmetry directions of momentum space \cite{Bohnen}. The agreement between the experimental results and the DFT data is very reasonable, considering that the calculations were performed for fully oxygenated YBa$_2$Cu$_3$O$_7$. The low-energy IXS intensity is dominated by a transverse acoustic (TA) phonon and an optical mode whose dispersions are well described by DFT. The constant-$\bf q$ profiles discussed below therefore exhibit a two-peak structure.

The temperature evolution of the phonon profiles in the vicinity of ${\bf q}_{CDW}$ is displayed in Figure 3. The full IXS spectrum \cite{SOM} is composed of Stokes and anti-Stokes components due to phonon creation and annihilation. As expected based on the detailed-balance theorem, the ratio between Stokes and anti-Stokes intensities everywhere in momentum space was observed to be perfectly described by the Bose thermal excitation factor \cite{SOM}. We note that this is in contrast to a recent report of deviations from detailed balance in a different cuprate \cite{Bonnoit}. The temperature dependent spectra of Fig. 3 D-F were therefore normalized by dividing out the Bose factor. The spectra were fitted to standard damped harmonic oscillator profiles convoluted with the experimental resolution function whose energy and in-plane momentum resolution were set to 3 meV and 0.015 reciprocal lattice units (full width at half maximum, FWHM), respectively. Figure 4 shows the results of these fits. Whereas the phonon dispersions at $T = 150$ K are identical to those at room temperature (Fig. 2) within the experimental error, and phonon linewidths are limited by the instrumental energy resolution, marked anomalies are observed at lower temperatures.

We first focus on the behavior in the superconducting state, where the dispersions of both phonons exhibit pronounced dips in the vicinity of ${\bf q} = {\bf q}_{CDW}$ (Figs. 3A, 4A and 4B). As these anomalies are restricted to a very narrow ${\bf q}$-range around ${\bf q}_{CDW}$, they were apparently not recognized in prior work on the lattice dynamics of YBa$_2$Cu$_3$O$_{6+x}$. Note, however, that anomalous phonon softening and broadening were reported for higher-energy Cu-O bond-bending and bond-stretching phonons at comparable wave vectors. These anomalies can now be clearly associated with the CDW correlations reported in the recent x-ray scattering experiments \cite{Ghiringhelli,Chang,Achkar,Blackburn,Blanco-Canosa}.

Similar soft phonon modes have been observed in other compounds with quasi-one-dimensional (quasi-1D) \cite{Renker,Hoesch} and quasi-2D \cite{Weber,Leroux} electronic structure that either exhibit CDW order \cite{Renker,Hoesch,Weber}, or are on the verge of a CDW transition \cite{Leroux}. Remarkably, the sharpness of the dispersion anomaly in YBa$_2$Cu$_3$O$_{6.6}$ closely resembles those associated with ``Kohn anomalies'' in quasi-1D CDW compounds with strongly nested Fermi surfaces \cite{Renker,Hoesch}. In contrast, the phonon anomalies associated with quasi-2D CDWs are considerably broader \cite{Weber,Leroux}, reflecting the less pronounced nesting of the associated Fermi surfaces. Model calculations for cuprates (which also exhibit quasi-2D Fermi surfaces) in the vicinity of CDW transitions yield Kohn anomalies whose widths are comparable to those of other quasi-2D CDW compounds, but much broader than those observed experimentally \cite{DiCastro}. However, previous scanning tunneling spectroscopy (STS) experiments have shown that the low-energy electronic transitions in the superconducting state of Bi-based cuprates are dominated by a small set of sharply defined wave vectors, as a consequence of the nesting properties of the $d$-wave gap function \cite{McElroy,Howald,Kohsaka,Wise,Parker}. A detailed analysis of IXS and STS data on the same material is required to quantitatively assess the correspondence between the Kohn anomalies shown in Fig. 4A and the ``quasiparticle interference patterns'' observed by STS. The same applies to recent photoemission data \cite{Vishik,Johnston} on low-energy ``kinks'' in the band dispersions of Bi$_2$Sr$_2$CaCu$_2$O$_{8+\delta}$, whose energies are in good agreement with those of our Kohn anomalies. We also note that the recovery signals in recent pump-probe experiments on underdoped YBa$_2$Cu$_3$O$_{6+x}$ \cite{Orenstein} and La$_{1.9}$Sr$_{0.1}$CuO$_4$  \cite{Torchinski} exhibits oscillations with frequencies that closely agree with those of the anomalous acoustic phonons we have observed.

We now turn to the behavior at higher temperatures, which is even more surprising. Upon heating up to $T_c$, the phonon energy is weakly $T$-dependent (although a slight softening of the acoustic phonon at ${\bf q} = {\bf q}_{CDW}$ is noticeable, Fig. 4B), and the linewidth remains resolution limited over the entire BZ, presumably because the maximum of the superconducting gap exceeds the phonon energy. At $T_c$, however, the frequency of the TA phonon at ${\bf q}_{CDW}$ abruptly jumps by about 15\% to its normal-state value; the hardening of the optical mode even exceeds $20\%$. At the same time, the phonon linewidths become extremely large in a narrow range around ${\bf q} = {\bf q}_{CDW}$, where the FWHM of the TA phonon at $T_c$ amounts to 3.5 meV, $\sim 40 \%$ of its energy. Upon further heating, the phonon gradually narrows, and the linewidth becomes resolution limited around $T \sim 150$ K, while the frequency is approximately $T$-independent.

On a qualitative level, the superconductivity-induced phonon anomaly we observed is in line with the generic behavior of low-energy phonons in superconductors, which are expected to broaden and harden when the energy gap collapses and low-energy electron-phonon decay channels open up upon heating above $T_c$. This behavior has indeed been confirmed both for acoustic phonons in conventional superconductors \cite{Axe,Aynajian} and for ${\bf q} =0$ optical modes in YBa$_2$Cu$_3$O$_{7}$ and other cuprates \cite{Friedl,Bakr}, but these effects do not exceed a few percent. To the best of our knowledge, the magnitude of the superconductivity-induced phonon anomalies in YBa$_2$Cu$_3$O$_{6.6}$ are by far the largest reported so far, undoubtedly as a consequence of the close competition between superconducting and CDW ground states in this material.

In order to further elucidate the nature of this competition, we now take a detailed look at the IXS data above $T_c$ (Fig. 5). The spectra are composed of an inelastic component due to Stokes and anti-Stokes scattering from phonons, as discussed above, and an elastic component centered at zero energy. The elastic line is $T$-independent and smoothly ${\bf Q}$-dependent over most of the BZ (Fig. 5A), and can thus be attributed to incoherent scattering from defects. For ${\bf q} = {\bf q}_{CDW}$, however, we observe an additional contribution to the elastic intensity (Figs. 5B,C). The $T$-dependent intensity and ${\bf q}$-width of this contribution are in excellent agreement with those inferred from the quasi-elastic scattering previously determined by resonant x-ray scattering experiments either without energy discrimination (inset in Fig. 5C) or with a much poorer resolution of $\sim 100$ meV \cite{Ghiringhelli}. In our high-resolution experiments, the CDW component of the elastic line remained energy-resolution-limited even after we enhance the energy resolution to $\sim 1.4$ meV FWHM \cite{SOM}. We can thus put an upper bound of $\sim 100$ $\mu$eV on the intrinsic energy width of the elastic component of the CDW signal. This implies that CDW domains with characteristic fluctuation energies below $\sim 100$ $\mu$eV and typical dimensions of $1-10$ nm (inferred from the momentum widths of the CDW peaks) are present in a wide temperature range both above and below $T_c$. Since this upper bound is still far above the characteristic energy scale of nuclear magnetic resonance (NMR) and nuclear quadrupole resonance (NQR), slow fluctuations of the CDW domains may preclude their observation by nuclear resonance techniques.

The elastic ``central peak'' in the IXS spectra of YBa$_2$Cu$_3$O$_{6.6}$ is closely similar to the behavior of other materials undergoing structural phase transitions, including insulating SrTiO$_3$ \cite{Cowley} and superconducting Nb$_3$Sn \cite{Axe}, which was attributed to defect-induced nucleation of domains of the low-temperature phase above the critical temperature. The nature of the lattice defects responsible for the ``central peak'' in these materials has remained undetermined, although oxygen vacancies were shown to play some role in SrTiO$_3$ \cite{Cowley}. In YBa$_2$Cu$_3$O$_{6+x}$, both local lattice distortions generated by oxygen defects in the CuO chains and extended defects such as dislocations may act as pinning centers for CDW nanodomains. The extremely large phonon linewidths in the normal state (Fig. 4C,D) can then be attributed to inhomogeneous broadening.

Since doping-induced lattice defects are present in all superconducting cuprates, the gradual onset of a spatially inhomogeneous CDW domain state with decreasing temperature may be a generic feature of the ``pseudogap'' regime in the cuprate phase diagram, although the temperature and doping dependence of the corresponding volume fractions may depend on the specific realization of lattice disorder in different materials. Whereas CDW nanodomains will surely contribute to the anomalous normal-state properties observed in this regime, their gradual nucleation explains the absence of thermodynamic singularities associated with CDW order, at least in the absence of a magnetic field \cite{LeBoeuf}. The persistence of this domain state over a much wider temperature range than corresponding phenomena in classical materials \cite{Cowley,Axe} probably reflects the strong competition between CDW correlations and superconductivity. In the presence of superconducting long-range order, the inhomogeneity is strongly reduced (Figs. 4 and 5). Conversely, thermodynamic singularities \cite{LeBoeuf} and NMR signals \cite{Wu} due to CDW long-range order have been reported in external magnetic fields strong enough to weaken or obliterate superconductivity. The fragility of the CDW nanodomain state in zero field and its competition with superconductivity explain the isotope effect on the superconducting penetration depth observed in the YBa$_2$Cu$_3$O$_{6+x}$ system \cite{Khasanov}. We also note the close analogy of our observations to the ``charge glass'' state previously identified by STS at low doping levels \cite{Kohsaka}, and to the nucleation of antiferromagnetic domains by spinless impurities studied by NMR \cite{Alloul} and neutron scattering \cite{Suchaneck}.

We end our discussion with some remarks about the implications of our results for the mechanism of high-temperature superconductivity. The electron-phonon interaction revealed by our IXS study appears strong enough to be a major contributor to at least some of the ``kinks'' observed in the dispersions of fermionic quasiparticles in the cuprates. Since it is very sharply concentrated in momentum space, however, its momentum-averaged strength seems insufficient to be a significant driving force for Cooper pair formation. Rather, the EPI favors a CDW instability that strongly competes with superconductivity and reduces the superconducting $T_c$ at moderate doping levels. We have confirmed these conclusions by repeating some of the IXS experiments on a fully oxygenated YBa$_2$Cu$_3$O$_{7}$ crystal with $T_c = 90$ K \cite{SOM}, where we found neither phonon anomalies characteristic of incipient CDW formation nor a central peak heralding CDW nanodomains.

\noindent \textit{Acknowledgement}. This work was performed at the ID28 beamline of the European Synchrotron Radiation Facility. We gratefully acknowledge L. Braicovich, M. Calandra, R. Comin, A. Damascelli, T. Devereaux, G. Khaliullin, and G. A. Sawatzky for insightful discussions.

\begin{thebibliography}{99}
\bibitem{SOM} Materials and methods are available as supporting materials.
\end{thebibliography}


\newpage

\bigskip

\begin{figure}
    \begin{center}
        \includegraphics[width=0.9\columnwidth]{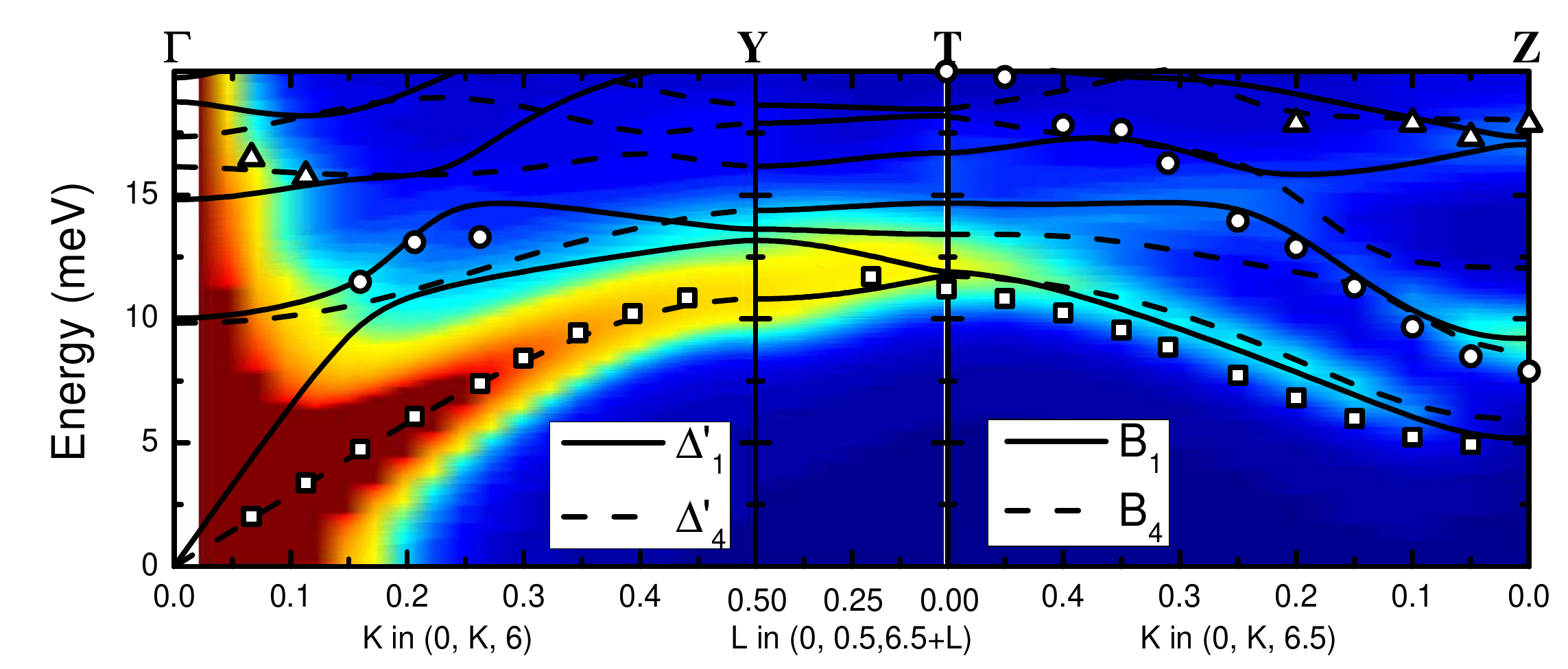}
    \end{center}
    \caption{Room temperature phonon dispersions along the high symmetry lines $\Gamma$Y, $YT$ and $TZ$ (Fig. 1). The white symbols represent the results of an analysis of IXS profiles, as described in the text. The lines and color map represent the results of DFT calculations for the phonon dispersions and IXS intensities, respectively \cite{SOM}. The symmetry of the phonon branches is indicated in the legend.}
    \label{fig:dispersion}
\end{figure}
\bigskip

\begin{figure}
    \begin{center}
        \includegraphics[width=0.9\columnwidth]{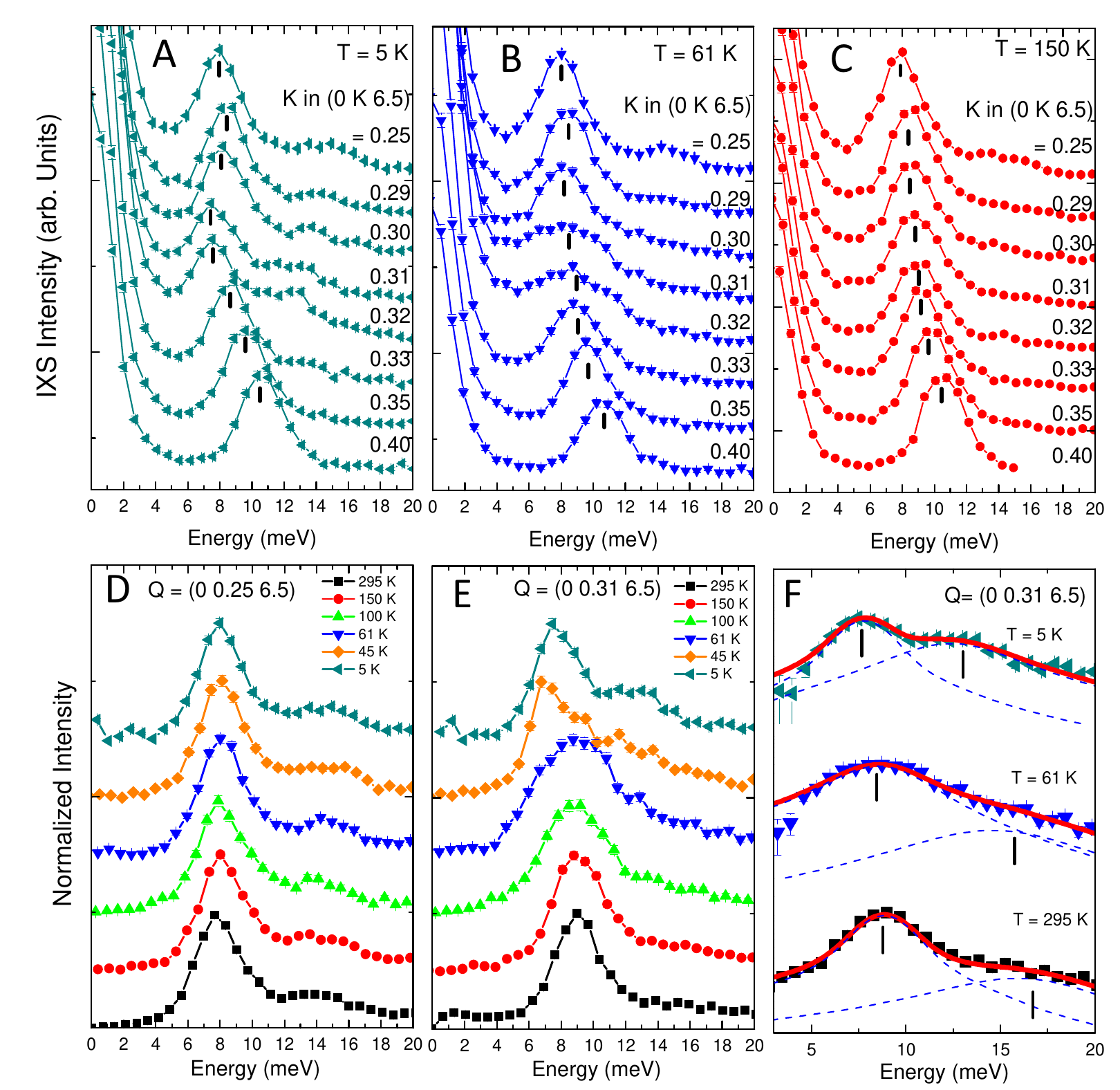}
    \end{center}
    \caption{Momentum dependence of the IXS spectra along the $ZT$ direction of reciprocal space at (A) $T = 5$ K, (B) $T=T_c=61$ K and (C) $T=150$ K. Black tick marks represent the phonon energy as extracted from fits to damped harmonic oscillator profiles \cite{SOM}. Panels (D) and (E) show the temperature dependence of the inelastic part of the IXS spectra at ${\bf q} = (0, 0.25, 6.5)$ and ${\bf q} = {\bf q_{CDW}} =(0, 0.31, 6.5)$, respectively. The intensities shown in these two panels were corrected for the Bose excitation factor, and then normalized to the intensity of the low-energy peak. (F) Details of the fits of the IXS spectra at ${\bf q_{CDW}}$ at 5, 61, and 295 K. A logarithmic scale has been used for clarity. The dashed lines and tick marks indicate the individual phonon profiles resulting from the fits and their maxima, respectively.}
    \label{fig:spectra_Tdep}
\end{figure}
\bigskip

\begin{figure}
    \begin{center}
        \includegraphics[width=0.9\columnwidth]{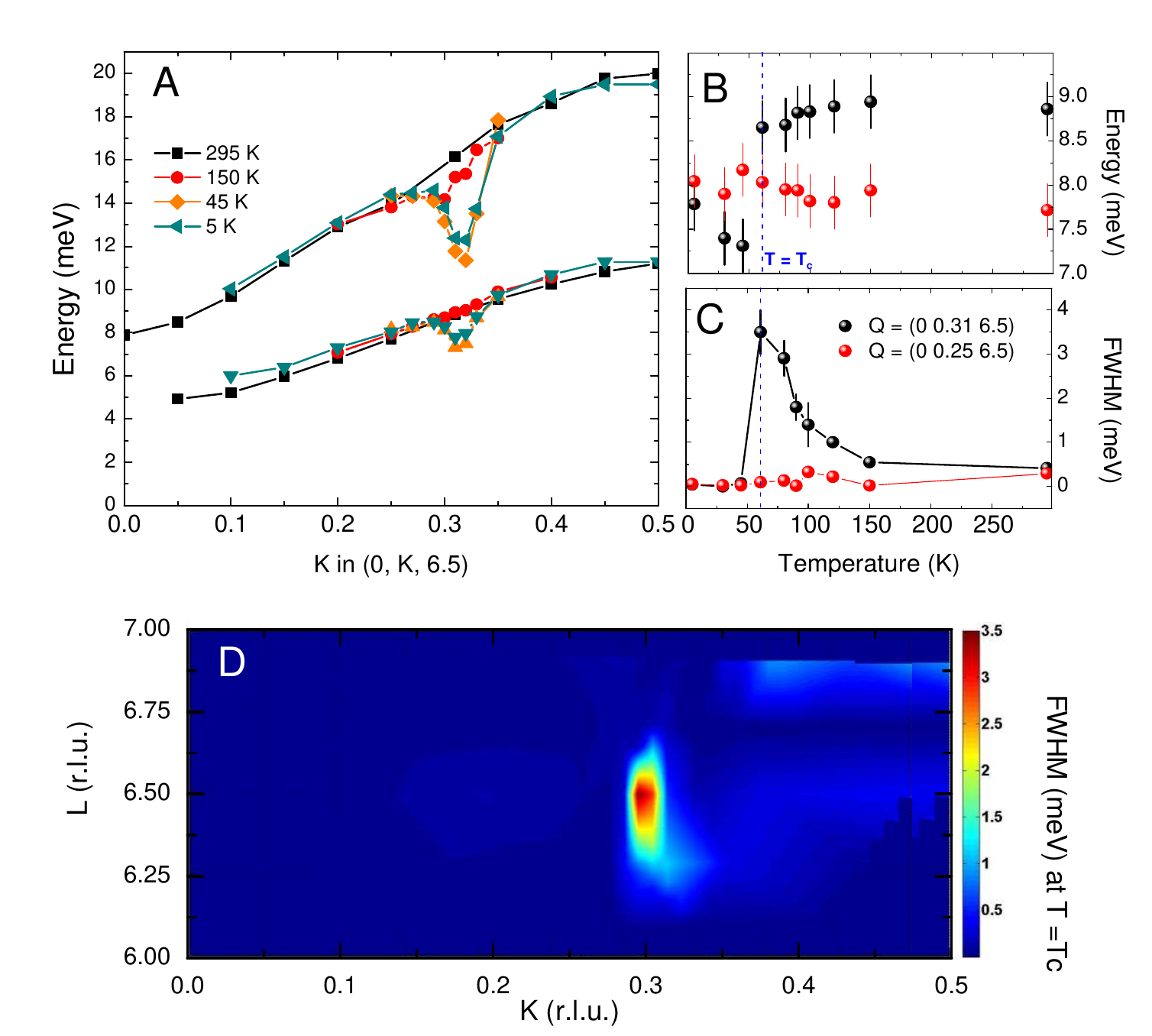}
    \end{center}
    \caption{ Temperature dependence of (A) the phonon dispersion and (B,C) the phonon lineshape parameters extracted from fits to the IXS spectra. (A) Dispersion of the two low-energy phonons in the $ZT$ direction at $T =$ 295, 150, 45, and 5 K. (B and C) Temperature dependence of the acoustical phonon energy and FWHM, respectively, at ${\bf Q} =(0, 0.25, 6.5)$ (red dots) and ${\bf Q} = (0, 0.31, 6.5)$ (black dots). In both panels, error bars represent the fit uncertainty. (D) Momentum dependence of the intrinsic FWHM of the TA phonon at $T = T_c$.}
    \label{fig:fitres}
\end{figure}

\begin{figure}
    \begin{center}
        \includegraphics[width=0.9\columnwidth]{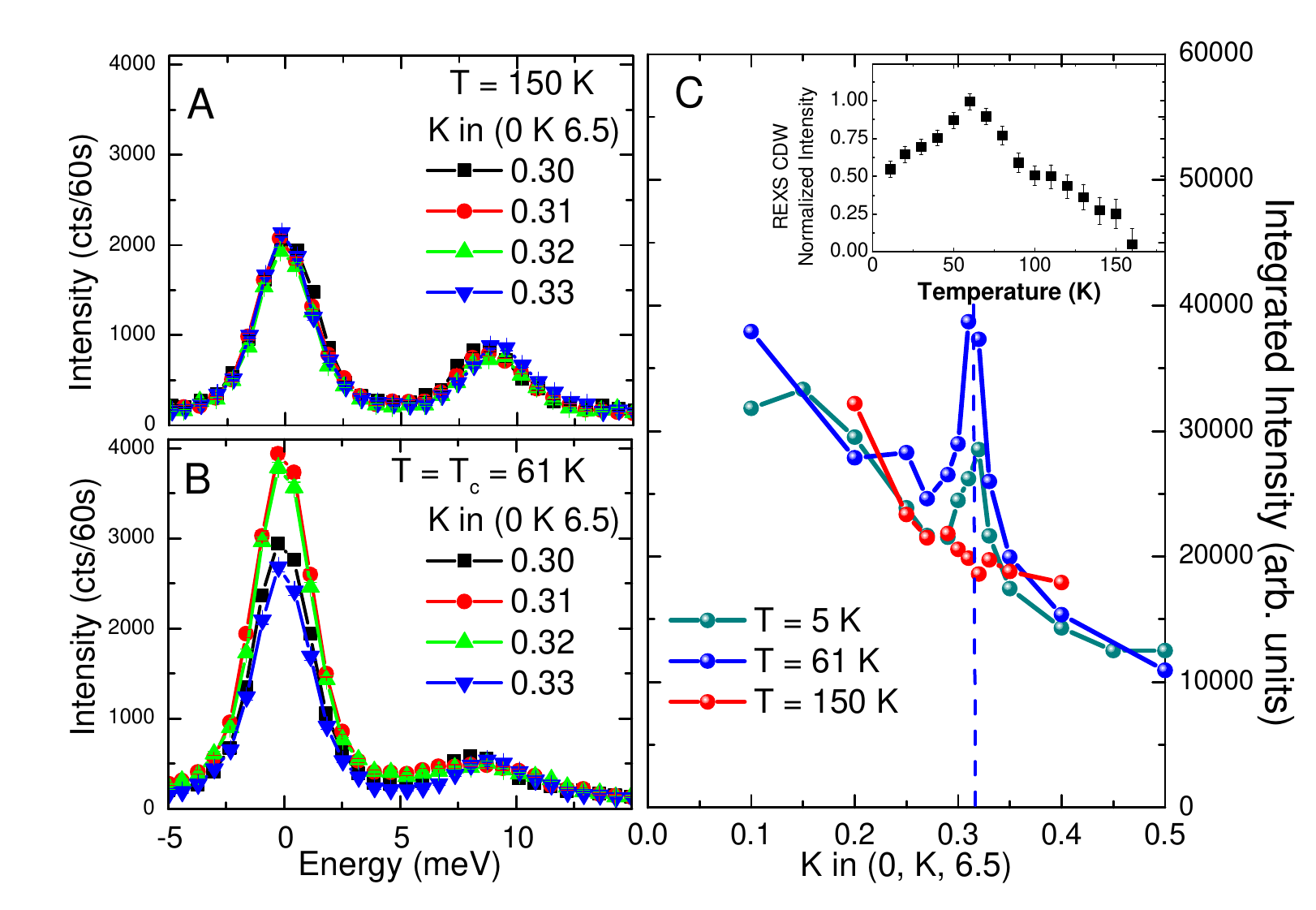}
    \end{center}
    \caption{Temperature dependence of the central peak. (A and B) IXS spectra in the vicinity of ${\bf q_{CDW}}$ at $T = 150$ K and $T = T_c = 61$ K, respectively, normalized for the Bose thermal excitation factor. Error bars represent the statistical error. (C) Momentum dependence of the intensity of the central peak at $T = 150$, 90, 61 and 5 K. The inset shows the temperature dependence of the CDW peak seen with resonant x-ray scattering \cite{Ghiringhelli}.}
    \label{fig:elastic}
\end{figure}
\bigskip

\newpage
\clearpage

\end{document}